\documentclass[aps,prd,
twocolumn,
floatfix,nofootinbib,groupedaddress,showpacs,showkeys]{revtex4-1}
\usepackage{graphicx}
\usepackage{amsmath}
\usepackage{latexsym}
\usepackage{here}
\usepackage{moreverb}
\usepackage{array}
\usepackage{dcolumn}% Align table columns on decimal point
\usepackage{bm}% bold math
\usepackage{epsfig}

\newcommand{\be}{\begin{equation}}
\newcommand{\ee}{\end{equation}}

\begin{document}

\title{Influence of the $Z-Z'$ mixing on the $Z'$ production cross section in the model-independent approach}

\author{A.~O.~Pevzner}
 \email{apevzner@omp.dp.ua}
\affiliation{Theoretical Physics Department, Oles Honchar Dnipro National University, 49010 Dnipro, Ukraine}

\date{\today}

\begin{abstract}
The new module {\it SMZp} is developed for the Monte-Carlo generator Sherpa that extends simulations of the Standard model (SM) processes with the Abelian $Z'$ boson in the model-independent approach. The special derived earlier relations between the $Z'$ couplings to the SM fields proper to the renormalizable theories are taken into account. Using this module, dependence of the $Z'$ production cross section on the $Z-Z'$ mixing angle $\theta_0$ in the Drell-Yan process is investigated within the range of 1~TeV~$\leq m_{Z'} \leq$~5~TeV. It is shown that if it is essential to keep the $Z'$ theory renormalizable, $\theta_0$ cannot be neglected as it is often done, even if it is small.
\end{abstract}


\maketitle

%%%%%%%%%%%%%%%%%%%%%%%%%%%%%%%%%%%%%%%%%%%%%%%%%
\section{Introduction}
%%%%%%%%%%%%%%%%%%%%%%%%%%%%%%%%%%%%%%%%%%%%%%%%%
Searching for the new particles is one of the main goals of the modern experiments in high energy physics. The hypothetical new heavy vector gauge $Z'$ boson is one of the most expected manifestations of new physics. A plenty of different scenarios exist where the $Z'$ appears \cite{Langacker:2008yv, Leike:1998wr, Salvioni:2009mt, Salvioni:2009jp}. The current model-dependent limits on $Z'$ mass  are of the order of 3.7--4.5~TeV \cite{Aaboud:2017buh,CMS:2015nhc}. Also, some recent constraints on the $Z'$ parameters can be found in \cite{Bobovnikov:2013kma, Andreev:2013ama, Gulov:2011yi, Gulov:2012zz, Gulov:2013dia, Pevzner:2015vwn}.

The $Z'$ search is typically considered in a framework of different $Z'$ models, such as $\mathrm{E_6}$, $\mathrm{LR}$, $\mathrm{ALR}$ etc. Each of them fixes the $Z'$ couplings to other particles, so that the $Z'$ mass stays the only free parameter which must be fitted from the experiment. These models are usually discussed as a natural benchmark for comparing the supposed $Z'$ signals in different experiments.

Despite the popularity of the model-dependent $Z'$ analysis, it may have some difficulties in the nearest future. The identification reach for the majority of the models is about the current estimated lower bound of the $Z'$ mass. This means it will be problematic to distinguish the  basic  $Z'$ model even if the $Z'$ is discovered experimentally. In such a  situation, a model-independent approach can be very prospective. In contrast to the model-dependent searches for the $Z'$ boson at the LHC
where only one free parameter exists (the mass of the
$Z'$) for a given $Z'$ model, in model-independent approach all fermion
coupling constants are considered as free parameters. Therefore, model-independent
approaches are prospective for estimating not only the mass but also $Z'$ couplings
to the SM particles. As a result, definite classes of the extended models could be restricted.
The obvious shortcoming of model-independent searches is a sufficiently large
amount of free parameters which must be fitted in experiments. However, it can be reduced if
some natural requirements or theoretical arguments are imposed. For example,
the considered Abelian $Z'$ boson assumes that the extended model is a renormalizable one.
This property results in the specific relations between couplings called in what follows
the renormalization group relations (RGR), that decreases the number of unknown parameters
significantly. The $Z$--$Z'$ mixing angle becomes also a self-consistent part of the parametric
space \cite{Gulov:2000eh}. RGR lead to the kinematical structure of the differential cross sections allowing picking out
uniquely the Abelian $Z'$ between other spin-1 neutral particles. In the Abelian class of $Z'$
models, the main $Z'$ properties and its couplings to the SM states can be completely
described in terms of three independent fermion couplings and the $Z'$ mass, $m_{Z'}$. Earlier, we applied this model-independent approach for the $Z'$ search in \cite{Pevzner:2015vwn, Pevzner:2017width, DirectSearch:2018}.

In any modern procedures dealing with the particle experiments, the Monte-Carlo (MC) simulations play a crucial role. They are extremely important for estimating the experiment background and predicting the valuable signals. So, it is a usual situation that any new theory or model is firstly being simulated using the MC methods. By default, some existing MC generators support simulations of the processes with participation of the $Z'$. At the same time no one among them includes a model-independent $Z'$ description. Within the present work, we implement a new model named {\it SMZp} for the Sherpa MC generator \cite{Sherpa}. This model extends the default Standard model file for Sherpa with the Abelian $Z'$ defined in our model-independent approach. The Sherpa generator is chosen due to convenience of developing the custom user-defined models. It allows introducing new fields and new interaction at a matrix element level. For example, PYTHIA \cite{PYTHIA}, one of the most well-known MC generators, can be extended with a new type of interactions by hardcoding the appropriate cross sections only.

The developed tool gives a possibility to make an advanced analysis of the processes where the $Z'$ can manifest itself. It is a tool which complements the package {\it DrellYan} for Wolfram Mathematica \cite{Pevzner:2017width} developed earlier. {\it DrellYan} allows analytical calculation of the Drell-Yan production cross section taking the $Z'$ contribution into account, while the {\it SMZp} model for Sherpa gives an opportunity for numerical computations of any possible processes with the $Z'$.

As an illustration of the developed module possibilities, we consider the process
\begin{equation} \label{DrellYan}
pp \to Z' \to l^+ l^- + X
\end{equation}
at the LHC and investigate how the $Z'$ production cross section $\sigma^{Z'}$ depends on the $Z-Z'$ mixing angle  $\theta_0$. In spite of the fact that the $Z-Z'$ mixing is often neglected because of its smallness (e.g., in the $\mathrm{B-L}$ model, in PYTHIA \cite{PYTHIA}), it occurs that dependence $\sigma^{Z'}(\theta_0)$ is quite important if the renormalization group relations mentioned above are taken into account.

The paper is organized as follows. In the next section, the information on the RGR necessary for what follows is given and the effective  Lagrangian describing interaction of the $Z'$ with the SM particles is adduced. In the Section 3, developing the {\it SMZp} module for Sherpa is described. Finally, in the Section 4, dependence of the $Z'$ production cross section on the $Z-Z'$ mixing angle is investigated for the Drell-Yan process. All the results are summarized in Discussion.

 \section{Effective Lagrangian and RGR relations}
 The effective low energy  Lagrangian
describing the interaction of the $Z$ and $Z'$ mass eigenstates
can be written as (see, e.g. \cite{Gulov:2000eh}):
\begin{multline}\label{ZZplagr}
{\mathcal{L}}_{Z\bar{f}f}=
Z_\mu\bar{f}\gamma^\mu[(v^\mathrm{SM}_{fZ}+\gamma^5
a^\mathrm{SM}_{fZ})\cos\theta_0 \\
 +(v_f+\gamma^5 a_f)\sin\theta_0]f,
\end{multline}
\begin{multline} \label{ZZ'plagr}
{\mathcal{L}}_{Z'\bar{f}f}=
Z'_\mu\bar{f}\gamma^\mu[(v_f+\gamma^5 a_f)\cos\theta_0 \\
- (v^\mathrm{SM}_{fZ}+\gamma^5a^\mathrm{SM}_{fZ})\sin\theta_0]f,
\end{multline}
where $f$ is an arbitrary SM fermion state; $v^\mathrm{SM}_{fZ}$,
$a^\mathrm{SM}_{fZ}$ are the SM axial-vector and vector couplings
of the $Z$-boson, $a_f$ and $v_f$ are the ones for the $Z'$, $\theta_0$ is the
$Z$--$Z'$ mixing angle. The definitions here are such that
\begin{multline} \label{couplings_SM}
v_f^{SM}=(T_{3,f}-2Q_f\hskip 2pt s_W^2)/(2s_Wc_W), \\
a_f^{SM}=-T_{3,f}/(2s_Wc_W).
\end{multline}
Within the considered formulation, the angle $\theta_0$ is determined by
the coupling $\tilde{Y}_\phi$ of fermions to the scalar field as
follows (see \cite{Gulov:2000eh} and Appendix B of
\cite{Pevzner:2015vwn} for details)
\begin{equation}\label{MixingAngle}
\theta_0 =
\frac{\tilde{g}\sin\theta_W\cos\theta_W}{\sqrt{4\pi\alpha_\mathrm{em}}}
\frac{M^2_Z}{M^2_{Z'}} \tilde{Y}_\phi
+O\left(\frac{M^4_Z}{M^4_{Z'}}\right),
\end{equation}
where $\theta_W$ is the SM Weinberg angle, $\tilde{g}$ is
$\tilde{U}(1)$ gauge coupling constant and $\alpha_\mathrm{em}$ is
the electromagnetic fine structure constant. Although $\theta_0$
is small quantity of order ($m^{2}_{Z}/m^{2}_{Z'}$), it
contributes to the $Z$ and $Z'$ bosons exchange amplitudes and
cannot be neglected.

As was shown in \cite{Gulov:2000eh,Gulov:2001sg}, if the extended
model is renormalizable and contains the SM as a subgroup, the
relations between the couplings hold:
\be \label{rgrav} v_f - a_f = v_{f^*} - a_{f^*}, \quad a_f =
T_{3f}\tilde{g}\,\tilde{Y}_\phi. \ee Here $f$ and $f^*$ are the
partners of the $SU(2)_L$ fermion doublet ($l^* = \nu_l, \nu^* =
l, q^*_u = q_d$ and $q^*_d = q_u$), $T_{3f}$ is the third
component of the weak isospin.
 These relations  are proper to the models of the Abelian $Z'$.
 They are just  as the correlations for the special  values of the
hypercharges $Y^R_f, Y^L_f, Y_\phi$  of the left-handed,
right-handed fermions, and scalars in the SM.  But now  these
parameters are some arbitrary numbers.

The correlations (\ref{rgrav})  have been already derived in two
different ways. The first one origins from the structure of the
  renormalization group operator and other one is founded on the
 requirement of the SM Yukawa term  invariance with
 respect to the additional $\tilde{U}(1)$ group.

From now on  we will assume that the axial-vector coupling $a_f$
is universal, so that
\be \label{a_universality} a \equiv a_e =
-a_{\nu_e} = a_d = -a_u =... \ee
Together with Eqs.~(\ref{MixingAngle}) and (\ref{rgrav}), it yields \be
\label{MixingAngle1} \theta_0 = -2a \frac{\sin\theta_W
\cos\theta_W}{\sqrt{4\pi\alpha_{em}}}\frac{M^2_Z}{M^2_{Z'}} +
O\left(\frac{M^4_Z}{M^4_{Z'}}\right). \ee

Eq.~(\ref{rgrav}) plays crucial role in what follows. First, due
to them the number of independent parameters is considerably
reduced. Second (but not less important) is influence on
kinematics of scattering processes. If the signal is observed, due
to these relations of Eq.~(\ref{rgrav}) one can  guarantee  that
exactly the Abelian $Z'$ state is observed.  Finally, it is important to notice that the relations (\ref{rgrav}) hold also in the Two-Higgs-Doublet Model (THDM). All this
makes the direct searching for the $Z'$ combined with (\ref{rgrav})
grounded and perspective.

\section{Development of the $Z'$ module for Sherpa 2.2}
As mentioned above, the Sherpa generator allows extending of its existing models with the custom ones. Several ways exist to implement a user-defined Sherpa model. For our purposes, we used the generic interface \cite{Hoe14c} to the UFO output \cite{Deg11} of FeynRules \cite{Chr08, Chr09}. It provides a suitable toolset for the model description in a FeynArts-like manner. Actually it gives a possibility to build a complete Lagrangian of a theory in a desirable parametrization.

The FeynRules $Z'$ model file was created basing on the SM model file which is built-in into Sherpa. The latter one requires several extensions to account for the $Z'$. First, we must introduce an additional (unphysical) $U(1)$ field, for example, $\tilde{B}^\mu$. This field should be a mixed state of the physical $Z$ and Abelian $Z'$ states. Then we complete the gauge sector of the SM with the $\tilde{B}$ kinetic term,
\begin{equation}
\mathcal{L}_{\mathrm{gauge}} = \mathcal{L}_{\mathrm{gauge}}^{\mathrm{SM}} + \frac{1}{2} m_{Z'}^2 \tilde{B}_{\mu} \tilde{B}^{\mu}.
\end{equation}
Finally, it is necessary to fix the $Z'$ couplings with the SM particles. Due to the relations (\ref{rgrav}), we can express all of them through $a$, $v_e$, $v_u$, and $m_{Z'}$. In FeynRules, the couplings are usually introduced in terms of a hypercharge, so that the covariant derivative describing the $Z'$ interaction with some particle $X$ will have a form
\begin{equation} \label{Y_def}
\mathcal{D}^\mu_{X} = \mathcal{D}^\mu_{X, \mathrm{SM}} - i \tilde{g} \tilde{Y}_X \tilde{B}^\mu.
\end{equation}
Here, $\tilde{Y}_X$ is the hypercharge of the field $X$ with respect to the $Z'$. So, we assign a new hypercharge $\tilde{Y}$ to each SM field. Using (\ref{ZZ'plagr}) and (\ref{Y_def}), it is easy to obtain its values for all the SM particles. They are summarized below in Table \ref{tbl:Hypercharges}.
\begin{table}[ht]
    \caption{The hypercharges $\tilde{Y}$ of the SM fields with respect to their interaction with the $Z'$}
    \label{tbl:Hypercharges}
    \centering
    \begin{tabular}{|l|c|}
        \hline
        Fields & $2 \tilde{g} \tilde{Y}$ \\ \hline\hline
        $\nu_{eL}$, $\nu_{\mu L}$, $\nu_{\tau L}$ & $v_e - a$ \\ \hline
		$e_{L}$, $\mu_{L}$, $\tau_{L}$ & $v_e - a$ \\ \hline
		$u_{L}$, $c_{L}$, $t_{L}$ & $v_e + a$ \\ \hline
		$d_{L}$, $s_{L}$, $b_{L}$ & $v_e + a$ \\ \hline
        $\nu_{eR}$, $\nu_{\mu R}$, $\nu_{\tau R}$ & $v_e + a$ \\ \hline
		$e_{R}$, $\mu_{R}$, $\tau_{R}$ & $v_e + a$ \\ \hline
		$u_{R}$, $c_{R}$, $t_{R}$ & $v_u - a$ \\ \hline
		$d_{R}$, $s_{R}$, $b_{R}$ & $v_u + 3a$ \\ \hline
		$\phi$ & $-2a$ \\ \hline
    \end{tabular}
\end{table}
After making these changes, the FeynRules model file is ready. Now it can be compiled for using with Sherpa. In the output module, the $Z'$ has an identifier 9900032. This module has five configurable parameters, namely the $Z'$ mass, $a$, $v_e$, $v_u$, and the $Z'$ decay width $\Gamma_{Z'}$. It is worth noting that actually $\Gamma_{Z'}$ is not an independent quantity, it is completely defined by the $Z'$ couplings if we ignore the exotic decay channels. Sometimes, however, it is useful to tune it manually. That is why this possibility is also left. Below, the fragment from the Sherpa configuration is shown that illustrates using of the created model.

\begin{verbatimtab}[4]
##############################################

MODEL SMZp;
# ...
block mass
	# ...
    9900032 4000.0   # m_{Z'} = 4 TeV
	# ...
block zprime
	1       -0.146   # a
	2        0.292   # v_e
	3        0.000   # v_u
# ...
decay 	9900032 40.0 # Gamma_{Z'} = 40 GeV
# ...

##############################################
\end{verbatimtab}

After the Sherpa module is ready, we should check its validity by simulating some processes where the $Z'$ takes part and comparing the results with the known ones. To do that, we calculate the inclusive cross sections of the Drell-Yan process, $pp \to \gamma/Z/Z' \to l^+l^- + X$ using Sherpa and compare it to the corresponding values obtained with our package {\it DrellYan} developed for Wolfram Mathematica earlier. In order to make sure of reliability of this comparison, we perform it for different $Z'$ masses in two regimes, "on-peak" ($M_1 < m_{Z'} < M_2$) and "off-peak" ($M_1, M_2 \ll m_{Z'}$), $M_1$ and $M_2$ are the bounds of the investigated invariant mass window ($M_1 < M < M_2$). All the calculations are carried out for $\sqrt{s} = 13$~TeV. The phase space cuts typical for the LHC experiments are applied, $|Y| < 2.4$ and $p_T^l > 20$~GeV, where $Y$ is a dilepton pair rapidity, $p_T^l$ is a lepton transverse momentum. The couplings of the $Z'$ to the SM leptons are set to those in the $Z'_\chi$ model.
\begin{table}[ht]
    \caption{Comparison of the $Z'$ contributions to the Drell-Yan production inclusive cross section calculated with Sherpa and {\it DrellYan}}
    \label{tbl:SherpaVsMathematica}
    \centering
    \begin{tabular}{|l|l|l|l|}
        \hline
        $m_{Z'}$, GeV & $M$, GeV & $\sigma_{Z'}$ (pb), Sherpa & $\sigma_{Z'}$ (pb), {\it DrellYan} \\ \hline\hline
        $1000$ & $100-200$ & $-0.012$ & $-0.012$ \\ \hline
        $1000$ & $800-1200$ & $0.06$ & $0.06$ \\ \hline
        $2000$ & $200-400$ & $0.0009$ & $0.0008$ \\ \hline
        $2000$ & $1800-2200$ & $0.004$ & $0.004$ \\ \hline	
        $3000$ & $400-60$0 & $0.0004$ & $0.0003$ \\ \hline	
        $3000$ & $2800-3200$ & $0.0005$ & $0.0004$ \\ \hline	
    \end{tabular}
\end{table}
The table shows a correspondence between the cross sections evaluated with Sherpa and {\it DrellYan} package in a wide range of invariant masses and for different $m_{Z'}$. Let us also note that the negative values of $\sigma^{Z'}$ in the table should not confuse anyone. Here $\sigma^{Z'}$ is defined just as a $Z'$ contribution to the total Drell-Yan cross section,
\begin{equation}
\sigma^{Z'} \equiv \sigma^{\mathrm{SM}+Z'} - \sigma^{\mathrm{SM}}.
\end{equation}
Hence $\sigma^{Z'} < 0$ may appear due to the $Z-Z'$ interference terms in the total cross section and causes no contradictions.

\section{Dependence of the $Z'$ production cross section on the $Z-Z'$ mixing angle}
The $Z-Z'$ mixing is an important dynamical characteristic of the $Z'$ boson. The present estimates on the $Z-Z'$ mixing angle $\theta_0$ are usually put as $|\theta_0| < 10^{-3}-10^{-4}$. Recently, it has been estimated in the diboson channel \cite{Osland:2017ema, Pankov:2016hun, Andreev:2014pra} and in the dilepton channel \cite{DirectSearch:2018} from the CMS and ATLAS data. At the same time a lot of investigations concerning the $Z'$ search exist the $Z-Z'$ mixing is neglected due to its smallness. However, as mentioned in Section 2, in a renormalizable theory, $\theta_0$ is not an arbitrary parameter as it is connected with the universal axial-vector coupling $a$ and the $Z'$ mass (see Eq. (\ref{MixingAngle1})). Hence just forcing $\theta_0$ to be zero is irrelevant in this framework. So, it is of interest to study how the $Z-Z'$ production cross section depends on the mixing angle.

In Fig. \ref{fig:ThetaDependence} below, we plot the $\sigma_{Z'}(\theta_0)$ dependence for 1~TeV~$ \leq m_{Z'} \leq $~5~TeV. The $Z'$ cross section is integrated over $m_{Z'} - 3\Gamma_{Z'} < M < m_{Z'} + 3\Gamma_{Z'}$ with $\Gamma_{Z'} = 0.01 m_{Z'}$ and $|Y| < 2.4$. The $v_e$ and $v_u$ couplings are put to those in the $Z'_\chi$ model and the $a$ coupling is varied accordingly to (\ref{MixingAngle1}) so that $\theta_0$ scales from $10^{-5}$ to $10^{-3}$.

\begin{figure}[h!]
    \caption{Dependence of the $Z'$ production cross section on the $Z-Z'$ mixing angle at different $Z'$ masses}
    \centering
    \includegraphics[scale=0.7]{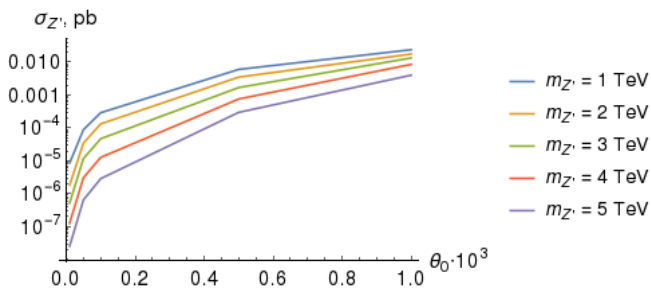}
    \label{fig:ThetaDependence}
\end{figure}

The figure above shows quite a strong dependence $\sigma^{Z'}(\theta_0)$. If $\theta_0$ changes by 2 orders, $\sigma^{Z'}$ is modified by about 4 orders. This is a direct consequent of (\ref{MixingAngle1}) and illustrates that in a renormalizable theory, $\theta_0$ cannot be excluded from consideration. If renormalizability requirement is not essential, this note can certainly be omitted.

\section{Summary and Discussion}
Within the present work, the module {\it SMZp} for the Monte-Carlo generator Sherpa was developed. It extends the default Sherpa functionality and allows the MC simulations with participation of the Abelian $Z'$ boson. The developed module accounts for the special relations (\ref{rgrav}) between the $Z'$ parameters proper to renormalizable theories. Therefore, the $Z'$ is parametrized by its mass $m_{Z'}$, universal axial-vector coupling $a$, vector coupling to the SM electron $v_e$, vector coupling to the SM up quark $v_u$, and decay width $\Gamma_{Z'}$. Generally speaking, if it is supposed that the $Z'$ does not interact with any particles except the SM ones, $\Gamma_{Z'}$  is completely defined by $a$, $v_e$, $v_u$, and $m_{Z'}$. At the same time sometimes it is necessary to set it manually. For example, it may be useful when dependence of some quantities on $\Gamma_{Z'}$ is investigated. Thus $\Gamma_{Z'}$ is also designed as an independent parameter of the {\it SMZp} module.

So, the Monte-Carlo simulations of the arbitrary processes with the $Z'$ described in the model-independent approach have become possible for the first time.

By using the developed tool, dependence on the $Z'$ production cross section on the $Z-Z'$ mixing angle $\theta_0$ was investigated in the Drell-Yan process $pp \to Z' \to l^+ l^- + X$ for 1~TeV~$\leq m_{Z'} \leq$~5~TeV. The present estimates on $\theta_0$ are about $|\theta_0| < 10^{-3} - 10^{-4}$, so it is often neglected. However, due to (\ref{MixingAngle1}), $\theta_0$ is not an arbitrary number, it is connected with $a$ and $m_{Z'}$. On this evidence, it is useful to understand how strongly the $Z'$ production cross section $\sigma^{Z'}$ depends on $\theta_0$ taking the mentioned relations into account. It turned out that varying $a$ in such a range that $\theta_0$ runs from $10^{-5}$ to $10^{-3}$ with keeping all other $Z'$ parameters constant increases $\sigma^{Z'}$ by $\sim 4$~orders. This fact is a direct consequence of the relation (\ref{MixingAngle1}). So, in a renormalizable theory, the $Z-Z'$ mixing cannot be neglected even if it is small.

In conclusion, we studied  Abelian $Z'$ bosons in the model-independent approach. In this framework, the $Z'$ can be parametrized only by the $Z'$ mass and three fermion coupling constants. We developed the {\it SMZp} module for the Shepra Monte-Carlo generator that extends the Standard model processes with the $Z'$ in our model-independent description. Using this tool, we investigated dependence of the $Z'$ production cross section on the $Z-Z'$ mixing angle in the Drell-Yan process. It was shown that if a $Z'$ theory is required to stay renormalizable, the $Z-Z'$ mixing cannot be neglected due to its connection with $a$ and $m_{Z'}$ parameters.

%%%%%%%%%%%%%%%%%%%%%%%%%%%%%%%%%%%%%%%%%%%%%%%%%
\section*{Acknowledgments}
%%%%%%%%%%%%%%%%%%%%%%%%%%%%%%%%%%%%%%%%%%%%%%%%%
The author acknowledges the receipt of the grant from the Abdus Salam International Centre for Theoretical Physics, Trieste, Italy. Also, he is grateful to Prof.~V.~V.~Skalozub and Prof.~A.~A.~Pankov for the fruitful discussions.


\begin{thebibliography}{18}

%\cite{Langacker:2008yv}
\bibitem{Langacker:2008yv}
  P.~Langacker, The Physics of Heavy $Z^\prime$ Gauge Bosons,
  Rev.\ Mod.\ Phys.\  {\bf 81}, 1199 (2009).
%  doi:10.1103/RevModPhys.81.1199
%  [arXiv:0801.1345 [hep-ph]].
  %%CITATION = doi:10.1103/RevModPhys.81.1199;%%
  %714 citations counted in INSPIRE as of 18 Jun 2016

%\cite{Leike:1998wr}
\bibitem{Leike:1998wr}
  A.~Leike,
  %``The Phenomenology of extra neutral gauge bosons,''
  Phys.\ Rept.\  {\bf 317}, 143 (1999).
  %doi:10.1016/S0370-1573(98)00133-1
  %[hep-ph/9805494].
  
%\cite{Salvioni:2009mt}
\bibitem{Salvioni:2009mt}
  E.~Salvioni, G.~Villadoro and F.~Zwirner,
  %``Minimal Z-prime models: Present bounds and early LHC reach,''
  JHEP {\bf 0911}, 068 (2009).

%\cite{Salvioni:2009jp}
\bibitem{Salvioni:2009jp}
  E.~Salvioni, A.~Strumia, G.~Villadoro and F.~Zwirner,
  %``Non-universal minimal Z' models: present bounds and early LHC reach,''
  JHEP {\bf 1003}, 010 (2010).

%\cite{Aaboud:2017buh}
\bibitem{Aaboud:2017buh}
  M.~Aaboud {\it et al.} [ATLAS Collaboration],
  %``Search for new high-mass phenomena in the dilepton final state using
  %36.1 fb$^{-1}$ of proton-proton collision data at $\sqrt{s}$ = 13 TeV with
  %the ATLAS detector,''
  arXiv:1707.02424 [hep-ex].

%\cite{CMS:2015nhc}
\bibitem{CMS:2015nhc}
  The CMS Collaboration [CMS Collaboration],
  %``Search for a Narrow Resonance Produced in 13 TeV pp Collisions Decaying to Electron Pair or Muon Pair Final States,''
  CMS-PAS-EXO-15-005.
  %%CITATION = CMS-PAS-EXO-15-005;%%

%\cite{Bobovnikov:2013kma}
\bibitem{Bobovnikov:2013kma}
  I.~D.~Bobovnikov and A.~A.~Pankov,
  %``Constraints on Extra Neutral Gauge Bosons from $W^+W^-$ Production at the ILC,''
  Nonlin.\ Phenom.\ Complex Syst.\  {\bf 16}, 382 (2013).

%\cite{Andreev:2013ama}
\bibitem{Andreev:2013ama}
  V.~V.~Andreev and A.~A.~Pankov,
  %``Distinguishing Indirect Signatures of New Physics at the International Linear Collider: Z' Versus Anomalous Gauge Couplings,''
  Nonlin.\ Phenom.\ Complex Syst.\  {\bf 16}, 51 (2013).

%\cite{Gulov:2011yi}
\bibitem{Gulov:2011yi}
  A.~Gulov and A.~Kozhushko,
  %``Model-independent estimates for the Abelian Z' boson at modern hadron colliders,''}
  Int.\ J.\ Mod.\ Phys.\ A {\bf 26}, 4083 (2011).

%\cite{Gulov:2012zz}
\bibitem{Gulov:2012zz}
  A.~V.~Gulov, A.~A.~Kozhushko, V.~V.~Skalozub, A.~A.~Pankov and A.~V.~Tsytrinov,
  %``Model-independent Z-prime searches at modern colliders,''
  Prob.\ Atomic Sci.\ Technol.\  {\bf 2012N1}, 48 (2012).

%\cite{Gulov:2013dia}
\bibitem{Gulov:2013dia}
  A.~Gulov and A.~Kozhushko,
  %``Estimates for the Abelian $Z'$ Couplings from the LHC Data,''
  Int.\ J.\ Mod.\ Phys.\ A {\bf 29}, 1450001 (2014).
  
%\cite{Pevzner:2015vwn}
\bibitem{Pevzner:2015vwn}
  A.~Pevzner and V.~Skalozub,
  %``Estimates of $Z'$ couplings within data on the $A_{FB}$ for Drell-Yan process at the LHC at $\sqrt s =$ 7 TeV and 8 TeV,''
  Phys.\ Rev.\ D {\bf 94}, no. 1, 015029 (2016).  

%\cite{Gulov:2000eh}
\bibitem{Gulov:2000eh}
  A.~V.~Gulov and V.~V.~Skalozub,
  %``Renormalizability and model independent description of Z-prime signals at low-energies,''
  Eur.\ Phys.\ J.\ C {\bf 17}, 685 (2000).  

%\cite{Gulov:2001sg}
\bibitem{Gulov:2001sg}
  A.~V.~Gulov and V.~V.~Skalozub,
  %``Renormalizability and searching for the Abelian Z' boson in four-fermion processes,''
  Int.\ J.\ Mod.\ Phys.\ A {\bf 16}, 179 (2001).

%\cite{Pevzner:2017width}
\bibitem{Pevzner:2017width}
  A.~Pevzner,
  %``Estimate of the Abelian Z' boson decay width''
  Visnik Dniprovs'kogo universitetu. Seria Fizika, radioelektronika, {\bf 7} (2017),
  arXiv:1712.03144v3 [hep-ph].

%\cite{DirectSearch:2018}
\bibitem{DirectSearch:2018}
  A.~V.~Gulov, A.~A.~Pankov, A.~O.~Pevzner, and V.~V.~Skalozub,
  Model-independent constraints on the Abelian $Z'$~couplings within the ATLAS data on the dilepton production processes at $\sqrt{s} =$~13~TeV,
  Nonlin.\ Phenom.\ Complex Syst. (in print) (2018).
  
%\cite{Sherpa}
\bibitem{Sherpa}
  T. Gleisberg, S. Hoeche, F. Krauss, M. Schoenherr, S. Schumann, F. Siegert, and J. Winter,
  %``Event generation with SHERPA 1.1``
  JHEP {\bf 02}, 007 (2009),
  arXiv:0811.4622 [hep-ph].

%\cite{PYTHIA}
\bibitem{PYTHIA}
  T. Sjostrand, S. Mrenna, P. Skands,
  %``A Brief Introduction to PYTHIA 8.1``
  Comput. Phys. Commun. {\bf 178}, 852 (2008),
  arXiv:0710.3820 [hep-ph].
 
%\cite{Hoe14c}
\bibitem{Hoe14c}
  S.~Hoeche, S.~Kuttimalai, S.~Schumann, and F. Siegert,
  %``Beyond Standard Model calculations with Sherpa''
  Eur. Phys. J. C {\bf 75}, 135 (2015),
  arXiv:1412.6478 [hep-ph].
  
%\cite{Deg11}
\bibitem{Deg11}
  C.~Degrande, C. Duhr, B.~Fuks, D.~Grellscheid, O. Mattelaer, and T.~Reiter,
  %``UFO - The Universal FeynRules Output''
  Comput. Phys. Commun. {\bf 138}, 1201 (2012),
  arXiv:1108.2040 [hep-ph].

%\cite{Chr08}
\bibitem{Chr08}
  N.~D.~Christensen and C.~Duhr,
  %``FeynRules - Feynman rules made easy''
  Comput. Phys. Commun. {\bf 180}, 1614 (2009),
  arXiv:0806.4194 [hep-ph].

%\cite{Chr09}
\bibitem{Chr09}
  N. D. Christensen, P. de Aquino, C.~Degrande, C.~Duhr, B.~Fuks, M.~Herquet, F. Maltoni, and S.~Schumann,
  %``A comprehensive approach to new physics simulations''
  Eur. Phys. J. {\bf C} 71, 1541 (2011),
  arXiv:0906.2474 [hep-ph].

%\cite{Pankov:2016hun}
\bibitem{Pankov:2016hun}
  A.~A.~Pankov, A.~V.~Tsytrinov and V.~A.~Bednyakov,
  %``High Precision Determination of Z - Z' Mixing in Diboson Production
at the LHC,''
  Nonlin.\ Phenom.\ Complex Syst.\  {\bf 19}, no. 2, 196 (2016).

%\cite{Osland:2017ema}
\bibitem{Osland:2017ema}
  P.~Osland, A.~A.~Pankov and A.~V.~Tsytrinov,
  %``Probing $Z$-$Z'$ mixing with ATLAS and CMS resonant diboson production data at the LHC at $\sqrt{s}=13$ TeV,''
  Phys.\ Rev.\ D {\bf 96}, no. 5, 055040 (2017).

%\cite{Andreev:2014pra}
\bibitem{Andreev:2014pra}
  V.~V.~Andreev and A.~A.~Pankov,
  %``Z-Z' Mixing Effects at the Large Hadron Collider,''
  Nonlin.\ Phenom.\ Complex Syst.\  {\bf 17}, no. 4, 346 (2014).
  
\end{thebibliography}
\end{document}